\documentclass[final,3p,times,twocolumn]{elsarticle}

\usepackage{epsfig}

\usepackage{amssymb}
\usepackage{amsmath}

\usepackage{lineno}

\journal{NIM A Proceedings of RICAP 2011}

\begin{document}

\begin{frontmatter}



\title{Astrophysical neutrino results}


\author{Thomas K. Gaisser\\
{\small for the IceCube collaboration}}
\address{Bartol Research Institute and Dept. of Physics and Astronomy
University of Delaware, Newark, DE 19716, USA\\
E-mail: gaisser@bartol.udel.edu}

\begin{abstract}
This paper is a brief review of the status of the search for astrophysical
neutrinos of high energy.  Its emphasis is on the search for
a hard spectrum of neutrinos from the whole Northern sky above the steeply
falling background of
atmospheric neutrinos.  Current limits are so low that they
 are beginning to
constrain models of the origin of extragalactic cosmic rays.
Systematic effects stemming from incomplete
knowledge of the background of atmospheric neutrinos are discussed.

\end{abstract}


\end{frontmatter}
 \linenumbers


\section{Introduction}

\begin{figure}[t!]
\begin{center}
\epsfig{file=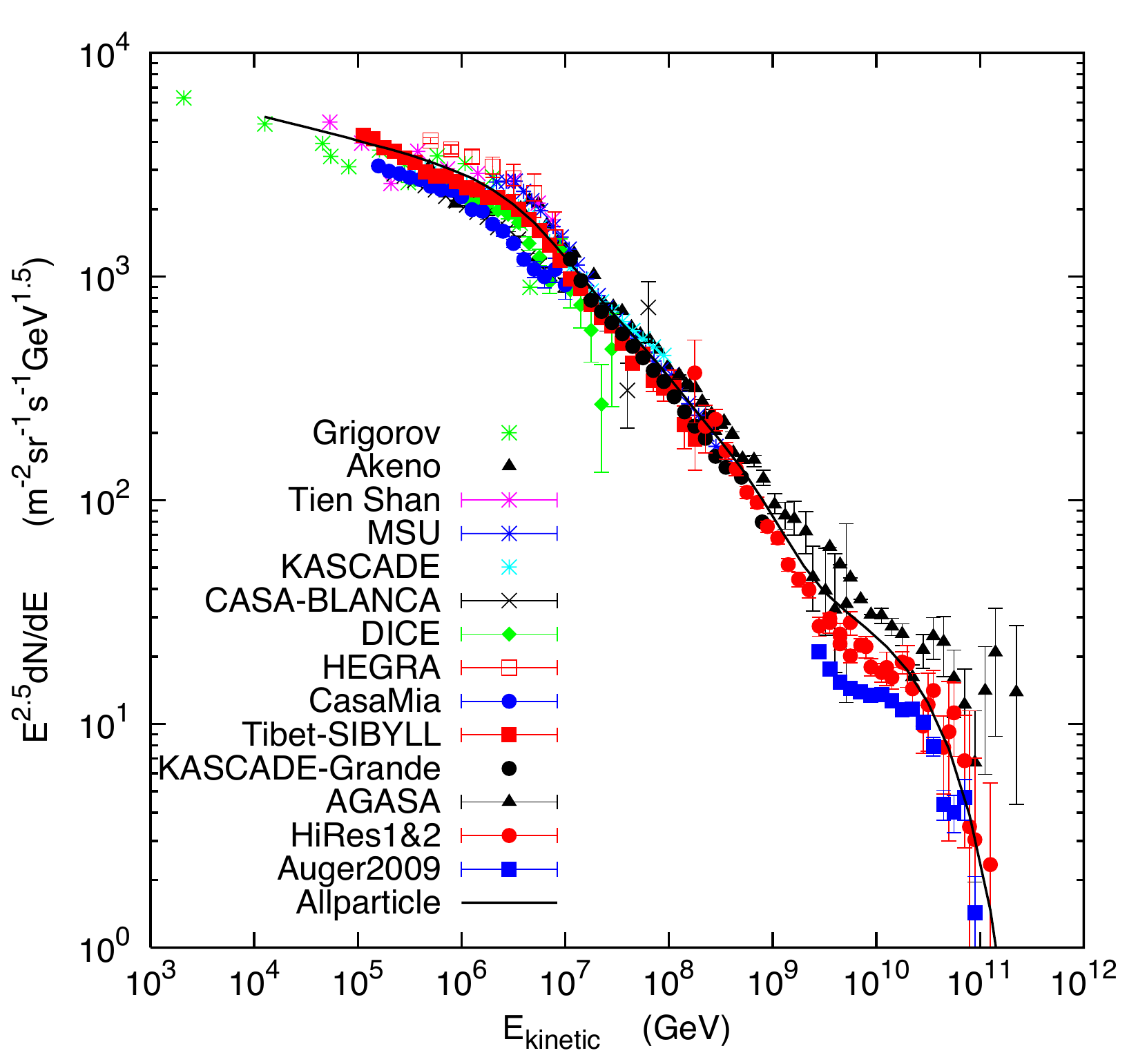,width=3.3in}
\caption{Compilation of the primary cosmic-ray spectrum
measured by air shower experiments~\cite{RPP}.
}
\label{fig1}
\end{center}
\end{figure}

One of the main goals of neutrino telescopes is to find neutrinos associated
with the sources of cosmic rays of ultra-high energy and thus to learn
about how the sources accelerate particles to $10^{20}$~eV.  
Figure~\ref{fig1} is a compilation of measurements of the
high energy cosmic-ray spectrum, which has three main features.  The
first is a knee
between $10^{15}$ and $10^{16}$~eV where the spectrum steepens from
$1.7$ integral spectral index to $2.1$.  The second is an ankle
between $10^{18}$ and $10^{19}$~eV above which the integral spectral index is $1.6$.  
Finally, there is a  
steepening above $5\times 10^{19}$~eV that is usually interpreted
as being the result of energy losses in the cosmic microwave background (CMB)
during propagation of particles over cosmic distances.
The cutoff above $10^{20}$~eV could
also just reflect the upper limiting energy of the cosmic accelerators.

The cosmic rays with energies above the
ankle are generally agreed to be of extra-galactic origin.
Somewhere at or below the ankle is the transition region
where the fluxes of particles from galactic sources are
comparable to those from extragalactic sources.
Exactly where the transition from galactic to extragalactic cosmic rays
occurs is an open question.

\begin{figure*}[thb]
\begin{center}
\epsfig{file=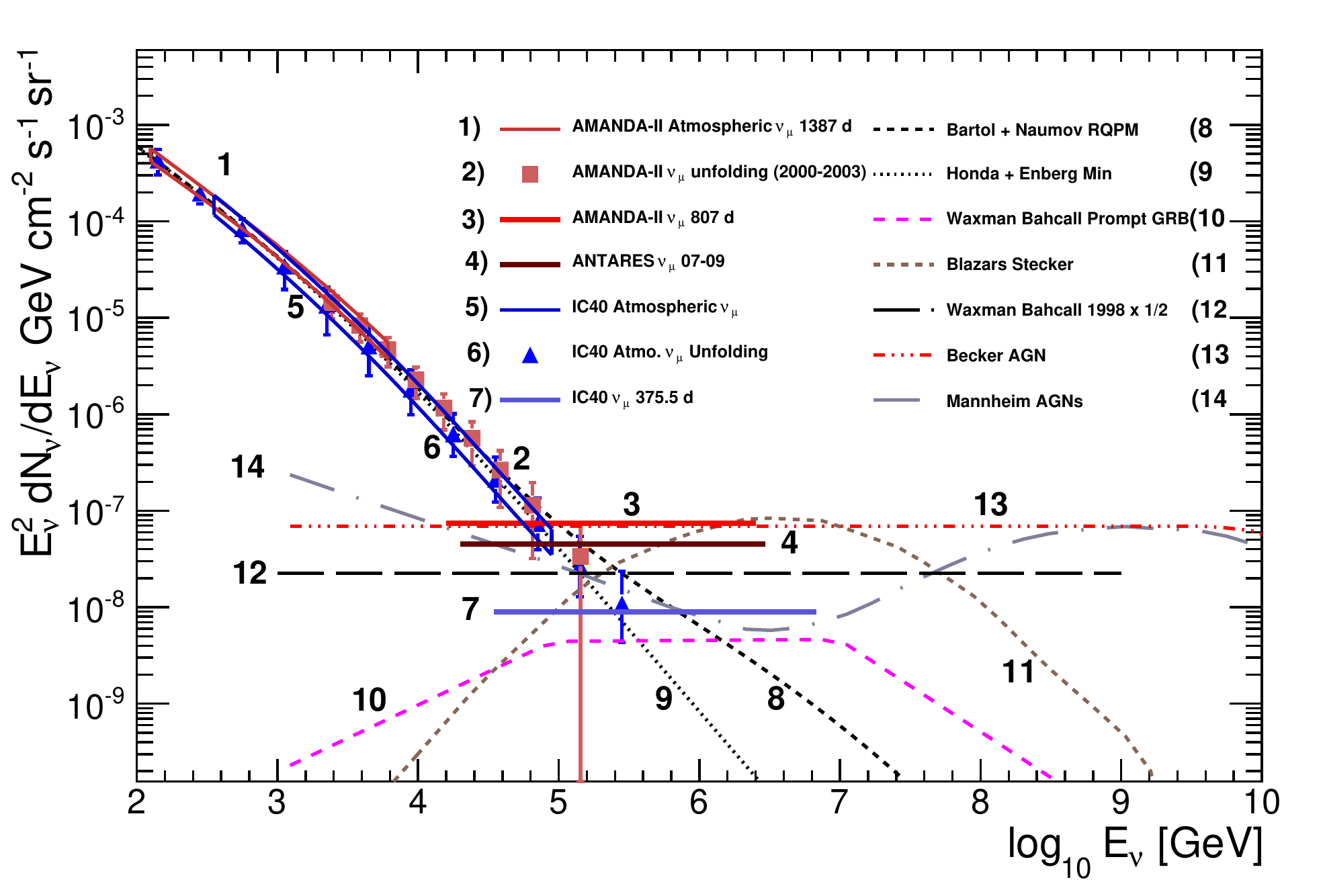,width=5.in}
\caption{Horizontal lines show limits on an $E^{-2}$ spectrum
of astrophysical muon neutrinos from AMANDA~\cite{AMANDA}, 
 Antares~\cite{Antares} and IceCube~\cite{IC40D}.  
The limits are shown along with measurements
of the flux of atmospheric muon neutrinos and anti-neutrinos.  The plot
is from Ref.~\cite{IC40D} where full references are given.
}
\label{fig2}
\end{center}
\end{figure*}

Two possibilities for extragalactic sources of cosmic rays
are often discussed.  One is the possibility
that the sources are active galactic nuclei (AGN), which is much debated
in connection with the Auger sky map of events
of the highest energy~\cite{Augerskymap}.  Another specific model
involves acceleration in the jets of gamma-ray bursts (GRB).
The model originally proposed in Ref.~\cite{WB-GRB} was extended
in Ref.~\cite{Guetta} to provide a predicted neutrino spectrum
for each burst based on measured features of its spectrum.
Recent limits from IceCube place constraints on both models.
The search with IceCube for neutrinos from identified GRBs was
covered in the main IceCube talk at this conference~\cite{Greg}.
This paper focuses on the search for an excess of 
astrophysical neutrinos from all directions that is expected
to emerge above the steeper background spectrum of atmospheric
neutrinos.

\section{Limits from IceCube with muon neutrinos}

The muon channel is the most favorable in terms of
event rate in the TeV range and above because the target
volume is enlarged by the charged current interactions of
neutrinos outside the detector
that produce muons that go through the detector.
The most sensitive analysis uses the Earth as a shield
against the downward background of cosmic-ray muons
by selecting horizontal and
upward moving events.  For energies in the TeV range and
above, stochastic energy losses by muons become
important, and the light produced increases in proportion
to the muon energy.  Simulations that incorporate the physics of neutrino 
interactions, of muon
energy loss and of ice properties are used to relate the
measured light to the energy of the muon in the detector
and thence to the energy of the neutrino.
This is done either by convolving an assumed neutrino 
spectrum with the sequence $\nu_\mu\rightarrow\mu\rightarrow$~observed light,
or by an unfolding procedure.  An important feature of the analysis
is that the distribution of $\nu_\mu$ energies that give rise to
a given signal in the detector is different for the steep
atmospheric neutrino spectrum from what it would be for a
hard spectrum of astrophysical neutrinos.

Upper limits on specific point sources of neutrinos in the Northern sky
from IceCube are currently less than $10^{-11}$~cm$^{-2}$s$^{-1}$TeV$^{-1}$.
These results were obtained in 2008-2009 when IceCube was half complete and
operating with 40 strings, each instrumented with 60 digital optical modules
at depths between 1450 and 2450 m in the ice at the South Pole.  The sensitivity
to point sources is approaching the level of $10^{-12}$~cm$^{-2}$s$^{-1}$TeV$^{-1}$~\cite{ptsrc} 
at which TeV gamma rays are seen from some blazars (e.g. Mrk 401~\cite{Daniel}).

It is important also to search for an excess
of astrophysical neutrinos from the whole sky at high energy above 
the steeply falling background of atmospheric neutrinos.  The
Universe is transparent to neutrinos, so the flux of neutrinos
from sources up to the Hubble radius could be large~\cite{Lipari}.
Limits from AMANDA~\cite{AMANDA}, Antares~\cite{Antares}
and IceCube~\cite{IC40D} are shown in Fig.~\ref{fig2}.
The current limit from the 40-string version of IceCube 
is now below the original Waxman-Bahcall bound~\cite{WB}.

This bound is an upper limit to the intensity of neutrinos which
holds if the neutrinos are produced in the same sources that produce
the extra-galactic cosmic rays.  The bound may be saturated if neutrinos
are produced from the accelerated particles that constitute the
ultra-high energy cosmic rays (UHECR).  Thus, as limits from IceCube
push below the Waxman-Bahcall bound, models of this type come into question.

\section{Atmospheric neutrino background ($\nu_\mu+\bar{\nu}_\mu$)}

Searching for an extraterrestrial flux of neutrinos 
below the Waxman-Bahcall limit is 
difficult because it is comparable
to the intensity of atmospheric neutrinos in an energy
region where this fundamental background is not well known.
The crossover between the current limit and the flux
of atmospheric muon neutrinos is between 200 and 1000 TeV,
depending on the level of neutrinos from decay of charmed hadrons
(giving prompt neutrinos). The level of prompt neutrino production is
highly uncertain, and the current IceCube limit appears
already to rule out the highest prediction
for charm~\cite{RQPM}.  

Moreover, the intensity of conventional
neutrinos from decay of $K^\pm$ and $\pi^\pm$ is itself uncertain.
Standard calculations of conventional atmospheric 
neutrinos~\cite{Bartol04,Honda06}
extend only to 10~TeV.  The atmospheric neutrino flux used
in the IceCube analysis~\cite{IC40D} is a simple power-law extrapolation
of the calculation of Ref.~\cite{Honda06}.  
Its normalization is treated as a free parameter in fitting
the data in Ref.~\cite{IC40D}, which is shown as a slightly
curved band that extends from 0.33 to 84~TeV in Fig.~\ref{fig2}.
The other experimental
results on the high-energy flux of atmospheric $\nu_\mu+\bar{\nu}_\mu$
in Fig.~\ref{fig2} are from AMANDA~\cite{AMANDA1,AMANDA2} 
and IceCube-40~\cite{Warren}.  All the atmospheric neutrino spectra
shown here are averaged over angle.  The unfolding analysis of Ref.~\cite{Warren}
extends to $E_\nu\approx 400$~TeV.  
The integral limit on astrophysical neutrinos shown for IceCube-40 in Fig.~\ref{fig2}
assumes a hard, $E^{-2}$ spectrum.  For this reason,
the bound applies at much higher neutrino
energies (35 TeV to 7 PeV) than the observed spectrum of atmospheric neutrinos.

Analyses of high-energy muon neutrinos with more recent versions
of IceCube are underway (with 59 strings in 2009-2010 and with
79 strings in 2010-2011).  The penultimate version with 79 strings
ran until May 20, 2011 in a configuration
that included the first 6 strings of a ``Deep Core" subarray~\cite{Ty}.
The full IceCube with 86 strings in the deep ice, including 8 Deep Core
strings, and 81 IceTop stations on the surface has been running
since then.  The IceCube-59 analysis~\cite{Anne} will remove
two simplifications made in previous analyses.  One is to take
account of the known zenith angle dependence of the atmospheric
neutrino flux, which increases toward the horizon.  This characteristic
dependence, known as the ``secant theta" effect, just reflects
the increased probability of decay compared to interaction of the
parent mesons for inclined trajectories in the upper atmosphere.
In contrast, both the astrophysical neutrinos and prompt neutrinos
(up to $\sim 10$~PeV) are isotropic.
Another improvement will involve taking account of the energy-dependence
of the atmospheric neutrino spectrum that follows from the knee and
other features of the primary cosmic-ray spectrum.  

Another source of uncertainty in the flux of atmospheric $\nu_\mu$ and
$\bar{\nu}_\mu$ at high energy arises from limited knowledge of
kaon production in the forward fragmentation region of hadronic
interactions in the atmospheric cascade~\cite{Agrawal}.  Kaons are the dominant
parent for neutrinos in the TeV energy range, and the process
of associated production ($p+{\rm air\;molecule}\rightarrow \Lambda + K^+ +$anything)
is particularly important.  The kaon contribution can be constrained,
and its contribution to the uncertainty reduced, by interpreting
the measurements of the muon charge ratio in the TeV energy range.  
The increase of $\mu^+/\mu^-$ in the TeV range
observed by MINOS~\cite{MINOScharge} and OPERA~\cite{OPERA} reflects the larger
charge ratio of kaons coupled with their increasing (though not dominant)
contribution to the muon flux in the TeV energy range.  

\section{Electron and tau neutrinos}

Although the muon channel is expected to have the highest event rate 
for both atmospheric and astrophysical neutrinos, it is also important
to look for electron and tau neutrinos.  These flavors are characterized
by the production of large, concentrated bursts of light (``cascades")
in the detector~\cite{Joanna}.
Neutrinos from distant sources have had time to
oscillate and are expected to arrive at Earth in equal (or at least
comparable) numbers in all flavors.  
The flux of atmospheric electron neutrinos is significantly
lower than atmospheric muon neutrinos until prompt neutrinos dominate.
In addition, the amount of light in the detector is
directly related to the total $\nu_e$ energy for charged current interactions.
For these reasons, the search for cascade-like events from interactions
of electron neutrinos inside the detector is important.  

Cascades at a level consistent
with charged current interactions of atmospheric electron neutrinos and neutral current
interactions of all flavors have been identified in the sub-TeV energy range
in Deep Core~\cite{Chang-Hyon}.  There are candidates for cascade
events at higher energy in IceCube-22~\cite{Joanna}
and in IceCube-40~\cite{Middell,Hickford}, but confirmation with IceCube-79
or with the full IceCube is needed to provide better containment for
this promising channel.

Since atmospheric tau neutrinos
are very rare, there is essentially no atmospheric background in this
channel.  Depending on the energy, tau neutrinos would show
up in the detector as two unresolved
cascades, as ``double bang" events~\cite{LearnedPakvasa} or as a cascade
associated with the track of a $\tau$-lepton, either entering or leaving
the instrumented volume~\cite{Reno}.

The cascade channels in IceCube have already been effective in contributing
to the limits at higher energy.  For example, in the search for
cosmogenic neutrinos with 
the 40-string version of IceCube~\cite{EHE}, the contribution to
the signal would be in the ratio $\nu_e:\nu_\mu:\nu_\tau\,=\,0.13:0.45:0.42$
assuming equal presence of the three flavors at Earth.  The energy flux
of neutrinos from energy loss to photo-pion production by protons of ultra-high
energy during propagation in the CMB typically
peaks around $10^{18}$ to $10^{19}$~eV when plotted as $E{\rm d}N_\nu/{\rm d}ln(E)$.
For a specific model~\cite{Ahlers} the expected number of events was
$\approx0.5$ in the IceCube-40 data sample.  No events passed the cuts, and an upper 
limit was set at $\le 3.6\times 10^{-8}$~GeV\,cm$^{-2}$s$^{-1}$sr$^{-1}$
for neutrinos of all flavors assuming an $E^{-2}$ differential spectrum.
The limit covers an energy range from 2 PeV to 6 EeV, overlapping
the energy ranges of cosmogenic neutrinos at high energy.

\noindent
{\bf Acknowledgment}: This work is supported in part by a grant from the U.S. National
Science Foundation, ANT-0856253.
\vfill\eject


\end{document}